\documentclass[12pt]{article}
\usepackage{graphicx}
\newcommand{\etal}{{\em et al\/}}
\newcommand{\vect}[1]{\mbox{${\bf #1}$}}

\newcommand{\seq}{{Schr\"odinger's equation}}
\newcommand{\beq}{\begin{equation}}
\newcommand{\eeq}{\end{equation}}

\newcommand{\hfm}[1]{\mbox{$\frac{#1}{2}^-$}}
\newcommand{\hf}{\mbox{$\frac{1}{2}$}}
\newcommand{\Bflam}{\mbox{\boldmath $\lambda$}}

\newcommand{\BfDel}{\mbox{\boldmath $\Delta$}}

\def\nuc#1#2{\relax\ifmmode{}^{#1}{\protect\text{#2}}\else${}^{#1}$#2\fi}

\begin{document}
\begin{center}
{\Large\bf Inverse scattering: applications to nuclear physics}\\[1 cm]
{R.S. Mackintosh$^{\dag}$, Department of Physical Sciences, The Open
University, Milton Keynes, MK7 6AA, U.K. }

\vskip 0.5 cm%

$^{\dag}$r.mackintosh@open.ac.uk,\\[3 mm]

{\bf Review commissioned for Scholarpedia}\\[1 cm]

Draft of \today\\

\end{center}

\vspace{5mm}

{\bf Abstract:}\, In what follows we  first  set the context
for inverse scattering in nuclear physics with a brief account
of inverse problems in general.  We then turn to inverse
scattering which involves the S-matrix, which connects the
interaction potential between two scattering particles with the
measured scattering cross section. The term `inverse' is a
reference to the fact that instead of determining the
scattering S-matrix from the interaction potential between the
scattering particles, we do the inverse. That is to say, we
calculate the interaction potential from the S-matrix. This
review explains how this can now be done reliably, but the
emphasis will be upon reasons why one should wish to do this,
with an account of some of the ways this can lead to
understanding concerning nuclear interactions.

\tableofcontents

\newpage

\section{General introduction: what are inverse problems?}
The subject of this review is a very specific part of the
research field `Inverse problems', a field of vast scope, with
dedicated journals including: Inverse Problems in Science and
Engineering (ISSN 1741-5977, print, and 1741-5985, online),
and,  Inverse Problems (ISSN 0266-5611, print, and 1361-6420,
online).

Inverse problems can be understood by contrast with the
corresponding direct problems, as some examples should make
clear:\begin{enumerate}
\item Given a distribution of electrical current within the
    brain, it is a straightforward \emph{direct} problem to calculate
    the magnetic fields outside the head; the \emph{inverse} problem
    of calculating the currents from measured magnetic fields
    is the much more difficult `biomagnetic inverse problem'.
\item Determining the size and location of a mass of iron ore
    from sensitive variations of gravity at the Earth's surface is much
    harder that the straightforward direct problem of
    calculating fluctuations in gravity at the surface due to known mass
    concentrations.
\item Various forms of tomography that  are central to modern
    medicine can be seen as inverse problems.
\end{enumerate}
Such problems have given rise to an active subdiscipline of
applied mathematics centred around integral equations; browsing
recent issues of the journals mentioned above will give a
flavour of the subject, and hint at its importance in modern
pure and applied science.

The subject of this article is much more restricted: the
application of inverse scattering in nuclear physics. The
nuclear inverse problem shares one key property with all of
those mentioned here: it is much harder and generally less
well-developed than the corresponding direct problem.  It
differs somewhat in that each of the other inverse problems is
widely accepted and plays a key role in the relevant
scientific, medical or commercial activity. The usefulness of
inverse methods in nuclear physics is less well-known; this
article will give examples of where it has been useful, and
maybe will inspire some new applications that this author has
not considered.

\section{Introduction to inverse scattering in nuclear physics}\label{intronp}
In this article, inverse scattering chiefly refers to \emph{the
determination of a local scattering potential that yields a given
set of S-matrix elements}. Although `true' nuclear interactions are
understood to be non-local, we discuss only the derivation of local
representations of the S-matrix. There always exists such a
representation and in view of the wide range of possible forms of
non-locality, the determination of a non-local potential from a
single set of S-matrix elements at a single energy will be
under-determined and we do not discuss this. We do remark, however,
that inversion can be the most natural way of determining a local
equivalent (in the sense of yielding the same S-matrix) of a
non-local potential. Inverse scattering  can also be extended to
include the determination of the interaction potential directly from
scattering observables (from this viewpoint, optical model fitting
is an elementary form of inverse scattering.)  The \emph{inverse} in
`inverse scattering' indicates a contrast with the (much easier)
\emph{direct} scattering problem in which the S-matrix, and thus the
scattering observables, are calculated from an interaction
potential. The physical context in which inverse and direct
scattering are discussed in this article is: the scattering of one
microscopic body from another in a model where the interaction
between the bodies is described by a potential and the natural
solution involves using this potential in the Schr\"odinger equation. In
accord with the title of this article, the microscopic bodies that
predominantly feature in this article are pairs of atomic nuclei,
with one of them often a nucleon.

This review sets out to do the following: \begin{enumerate}
\item Define the categories of nuclear inverse scattering cases,
specifying those that will be given fuller treatment in this review and
giving references for those that will not.
\item Present an overview of the various methods that have been
applied to the nuclear inverse scattering problem.
\item Present an account of a particular inversion procedure,
the iterative-perturbative (IP) inverse scattering algorithm, that has
had a wide range of applications. This will include specific examples
of what it can do.
\item Show `what inversion can do for you'. This takes the form of a
range of examples showing what kind of information or understanding
concerning nucleus-nucleus interactions can be obtained using inversion.
At the end, we leave it up to the imagination of the reader to extend that range.
\end{enumerate}
This review is specifically \emph{not} a comprehensive review of the
formal theory of inverse scattering.

\section{Definitions and notation}\label{def}
The \emph{direct} scattering problem, which is in the background to
this review, is the calculation of the scattering of interacting
microscopic particles, typically but not exclusively a nucleon and
an atomic nucleus. We assume that there is an interaction potential
between the interacting particles that can be substituted into the
time independent \seq\footnote{We do not consider wave packets nor
do we consider the justification for the stationary state treatment
of scattering.}.  This can be solved for the radial wave function
for specific values of the orbital angular momentum $l$ (we
generalize shortly to the case where the particle has spin) having
asymptotic form \beq  u_l(r) \to I_l(r) - S_l O_l(r) . \label{Sl}
\eeq Eqn.~\ref{Sl} defines the S-matrix $S_l$ for orbital angular
momentum quantum number $l$ for a spinless projectile; $I_l(r)$ and
$O_l(r)$ are the ingoing and outgoing radial solutions, Coulomb wave
functions, for the case where the projectile and target are both 
charged. The S-matrix is frequently expressed in terms of phase-shifts
$\delta_l$: $S_l= \exp{2 i \delta_l}$. For complex potentials,
$\delta_l$ is complex and $|S_l| \le 1$ to preserve unitarity.

For a spin-\hf\ projectile, we have \beq  u_{lj}(r) \to I_{l}(r) -
S_{lj} O_{l}(r)  \label{Slj}\eeq where $j = l\pm \hf$. The case of
spin-1 projectiles, such as deuterons,  can also be handled; this
involves the inversion of a coupled channel S-matrix: \beq
u^J_{l'l}(r) \to \delta_{l'l} I_{l'}(r) - S^J_{l'l} O_{l'}(r)
\label{SJll} \eeq where $J$ is the total angular momentum (assuming
spin zero target nucleus), $l$, $l' = J \pm 1$ and have the same
parity; for $l=l'=J$ there is no coupling.

The \emph{inverse} problem in these cases reverses the situation:
given $S_l$, $S_{lj}$ or $S^J_{l'l}$, what is the interaction
potential? We shall also touch on the inverse problem of
establishing $S_l$, $S_{lj}$ or $S^J_{l'l}$ from measured
observables.

{\bf General references:} The book by K. Chadan and P.C.
Sabatier~\cite{chadan} presents a comprehensive  account of inverse
scattering with an emphasis on the formal aspects.  We refer to this
as CS89. An old but still useful introduction to inversion is
Chapter 20 of the book by R.G. Newton~\cite{newton}.

{\bf Review article:} The review, `The application of inversion
to nuclear physics', by Kukulin and Mackintosh~\cite{kukmac}
reviews inverse scattering particularly as applied to nuclear
scattering, up to 2003 and provides a much more comprehensive
bibliography than the present one. We refer to it as KM04.

{\bf Conference proceedings:} Many articles discussing the
theory and application of inverse scattering, as well as
broader aspects of inversion, can be found in the conference
proceedings:~\cite{berlin,honnef,balaton}.

\section{Categories of inverse scattering}\label{categories}
In this review we shall refer to the following categories of inverse scattering:

{\bf 1. Fixed-$l$ inversion}\\
Given $S_l(E)$ for all energies $E$ at a fixed $l$, determine the
potential $V(r)$ that gives $S_l$. This is the classical inversion
problem solved by Gel'fand and Levitan and also Marchenko, see CS89. The term
`fixed-$l$ inversion' is misleading since it has been generalized to
include derivation of spin-orbit and tensor terms for specific fixed
$J$ and parity, see Sections~\ref{fixl} and~\ref{NN}.

{\bf 2. Fixed energy inversion}\\
Given $S_l$ for all $l$ at a specific energy $E$, fixed $E$
inversion determines $V(r)$ that reproduces $S_l$ at energy $E$. We
include under this heading inversion  $S_{lj} \to  V(r) + {\bf l
\cdot s}\ V_{\rm SO}(r)$ and also the generalization for spin-1
projectiles leading to the determination of a specific tensor
interaction from $S^J_{ll'}$ (where $(-1)^l= (-1)^{l'}$). Scattering
from target nuclei with spin can sometimes be treated by determining
independent interactions for different values of the channel spin.
Applications in which more general non-diagonal S-matrices are
inverted to non-diagonal potentials (coupled channel inversion) have
been discussed.

{\bf 3. Mixed case inversion}\\
At low energies, there may be too few active partial waves for
satisfactory inversion. If there exist sets of $S_l$ at closely
spaced energies, they can be inverted together to determine a
potential in `mixed case' inversion, having aspects of fixed-$l$ and
fixed-$E$ inversion. It can be viewed as incorporating information
from the local energy dependence of $S_l$. Mixed case inversion is
possible with the iterative-perturbative (IP) inversion procedure
that is presented in Section~\ref{IP}.

{\bf 4. Energy-dependent inversion}\\
Nuclear potentials are inherently energy dependent. Given $S_l$ for
a wider range of energies than is appropriate for mixed case
inversion, energy-dependent inversion determines a potential with an
appropriately parameterized energy dependence, and can be considered
a generalization of mixed case inversion.

{\bf 5. Variants of fixed energy, mixed case and energy-dependent inversion}\\
Scattering of identical bosons provides $S_l$ for just the even
values of $l$. Where there are sufficient active partial waves, this
situation can be handled straightforwardly by the IP method and
semi-classical (WKB) methods. Likewise, it is often straightforward
to obtain with IP inversion a parity-dependent potential in cases
where exchange processes (for example) require separate potentials
for odd-parity and even-parity partial waves (see Remark 4, below).

{\bf 6. Direct observable-to-potential inversion}\\
Using IP inversion, it is possible to combine in one algorithmic procedure
$S_l \to V(r)$ inversion together with a determination of $S_l$ from a fit to
data. This can be applied with any of types 2 to 5 above. (This is
distinct from the two-step inversion mentioned in Remark 2 below.)

\emph{Remark 1.} In principle, type 1 (fixed $l$) requires $S_l$ for
all energies and type 2 (fixed $E$) requires $S_l$ for all $l$, but
practical implementations have been developed. For fixed energy inversion,
this allows a potential to be defined out to a specific radius to be
determined from a set $S_l$ over an appropriate range of $l$. This
effectively puts a lower limit to the energy for which a potential
can be determined.

\emph{Remark 2.} In principle
all methods are subject to problems of non-uniqueness and errors,
though these problems can often be minimized in practice. Important
for this is the fact that practical inversion methods may allow the
inclusion of \emph{a priori} information, especially in cases
tending to be under-determined.

\emph{Remark 3.} In practical applications, the S-matrix to be
inverted generally comes from theory or from fits such as R-matrix
fits or effective range fits to measured observables over a range of
energies. There have also been applications in which $S_l$ have been
determined by fitting observables at a single energy with a direct
search. Such S-matrix fitting is also an inverse problem and the
technical and formal aspects of observable-to-$S_l$ inversion are to
be found elsewhere, e.g.\ CS89. We do mention some applications of
the resulting `two-step' phenomenology, which might be considered as
an alternative form of model-independent optical model (OM) fitting,
having certain advantages. This is true also of type 6, direct
observable-to-potential inversion.

\emph{Remark 4.} Methods for both fixed-$l$ and fixed-$E$
exist for including the energy of
bound states as input information for the inversion.

\emph{Remark 5.} A parity-dependent component (e.g. real or imaginary
central, real or imaginary spin-orbit) of a potential may be
written $V_{\rm W} + (-1)^l V_{\rm M}$ and, in the context of parity
dependence, we refer in what follows to $V_{\rm W}$ and $ V_{\rm M}$
as the Wigner and Majorana components. We reserve the term
`$l$-dependent' for other forms of partial wave dependence, never for
parity dependence.

\section{Alternative inversions}\label{alt}
As an alternative to determining the potential that reproduces the
S-matrix, one can determine the potential that reproduces the radial
wave function for a given partial wave. The trivially equivalent
local potential (TELP) of Franey and Ellis~\cite{fellis} is
necessarily $l$-dependent. However $l$-weighted TELPs, applicable
for all $l$, can be constructed (as in the CC code FRESCO~\cite{fresco})
and are widely used to represent dynamic polarization potentials,
DPPs, (discussed below). As suggested by, for example, the somewhat
arbitrary nature of the partial wave weighting, weighted  TELPs
cannot be expected to give the same potential as $S_l \to V(r)$
inversion and actual comparison~\cite{pangm} confirms that indeed
there are substantial differences. The consideration of such
alternative ways of defining a local potential can throw light on
the physics of local potential models of scattering, as discussed by
M. S. Hussein, \etal~\cite{hussein}.

As an alternative to the $l$-dependent TELP,  one can produce a
spatial representation of the potential $V(\vect{r})$ that
reproduces the elastic channel wave function $\psi(\vect{r})$
throughout the nucleus; note the vectorial dependence upon \vect{r}.
Such a representation illuminates the non-locality induced by
channel coupling, showing regions where flux leaves and then returns
to the elastic channel. The real and imaginary parts of the
`$\psi$-potentials' are determined, respectively, from the real and
imaginary parts of $\psi(\vect{r})^* \nabla^2 \psi(\vect{r})$ as
described in Ref.~\cite{npa494}, and for spin-\hf\ projectiles,
Ref.~\cite{prc49} and references therein. Ref.~\cite{npa542}
compares the very different wave functions within the nucleus for
$l$-dependent and $l$-independent potentials that have the same
asymptotic form, i.e.\ the same $S_l$.

\section{Methods for $S_l \to V(r)$ inversion}\label{proc}
A number of techniques for $S_l \to V(r)$ inversion have been put
forward and here we list the most significant with the emphasis on
the historically significant and those that have been widely
applied, leading to the understanding of nuclear interactions.

In principle, as has been pointed out by Chadan and
Sabatier~\cite{chadan}, the nuclear scattering inverse problem
is under-determined and hence subject to ambiguities. This is
more of a problem for formal methods that do not readily
permit the inclusion of prior information. In fact, it proves
not to be a problem in many practical cases, especially where
the sought-for potential is not too far from some known
potential, as, for example, when determining a dynamic
polarization potential. In other cases, it does matter, and the
inclusion of prior information in the overall inversion problem
has to be accepted as reasonable. It's a strength of certain
inversion procedures that this is possible with them. A
specific example will be given.

A possible consequence of under-determination is the occurrence of
rapid wiggles on the potentials that are determined. In effect,
potentials $V(r)$ and $V(r) + \tilde{V}(r)$, where $\tilde{V}(r)$,
a `null potential'~\cite{rsmnull}, is a function in the form of
a set of short wavelength oscillations, have exactly, or very nearly
the same $S_l$. This is particularly significant since genuine wavy
features do occur, e.g.\ as a consequence of underlying
$l$-dependence. Such genuine waviness must be distinguished from the
spurious. The IP method does afford means for such discrimination.

\subsection{Fixed-$l$ methods}\label{fixl}
The inversion formalism of Gel'fand, Levitan and Marchenko~\cite{chadan}
can be made to  yield a spin-orbit potential from
$S_{lj}$ and also (in coupled-channel form) yield a tensor
force when different $l$ values contribute to the S-matrix for
specific total angular momentum $J$ and parity $\pi$. The problems are:
(i) nuclear potentials are typically energy dependent but the method
relates $S_l$ to a single potential for the whole energy range, and (ii) a
very wide energy range is required to determine the potential.
Even for nucleon-nucleon scattering, where the method has been
applied, the pion threshold limits the energy range.

\subsection{Fixed-$E$ methods and extensions}\label{fixe}
Newton~\cite{newton,newton1}, starting from the fixed-$l$ formalism
of Gel'fand and Levitan, devised a restricted fixed-$E$ inversion
procedure that was further developed by Sabatier~\cite{chadan} and
others (see Section~\ref{NS}) into the Newton-Sabatier (NS)
inversion method. This formalism directly derives a potential from
S-matrix elements and is formally exact. The related method of
Lipperheide and Fiedeldey~\cite{lf,FL84} starts from a specific
parameterization of $S_l$.

Fixed-$E$ inversion methods for application at higher energies based
on the JWKB approximation and other semi-classical approximations
have been developed, see Kujawski~\cite{kujawski} and
others~\cite{SLW}. For applications see Section~\ref{WKB}.

The most widely applied inversion method is the
iterative-perturbative (IP) algorithm~\cite{mk82} based upon
the generally linear response of $S_l$ to changes in
$V(r)$~\cite{MacKob79}.  IP inversion has been extended to
handle mixed-case inversion, energy-dependent inversion,
spin-\hf\ inversion and some cases of spin-1 inversion leading
to a tensor interaction.

A number of other approaches to fixed-$E$ inversion are referenced in
the review~\cite{kukmac}. The subject of fixed-$E$ inversion is a topic
of on-going research, see  e.g.~\cite{pa2012}.

\subsubsection{Newton-Sabatier and related methods}\label{NS}
The formal Newton-Sabatier (NS) inversion method was developed
into a practical applicable method in the important work of
M\"unchow and Scheid~\cite{ms} (MS). Key aspects were the
matching of the radial range to the range of $l$-values for
which $S_l$ was provided, and the adoption of an
over-determined matrix algorithm.

For the later extension of the MS method to spin-\hf\ see Ref.~\cite{kukmac}.

Formal inversion methods of this kind simply translate a set of
$S_l$  values to a function $V(r)$ with the disadvantage that when,
for example, suspected `noise' in the input $S_l$ leads to
oscillatory features in $V(r)$, it is not possible to adjust the
precision required of the inversion to evaluate the physicality of
these features. It is also not straightforward to include prior
information concerning the potential; such information is useful in
difficult applications and for eliminating the effects of the
general under-determination of nuclear inverse
scattering~\cite{chadan}. Test inversions do appear to exhibit some
spurious oscillations that might be difficult to distinguish from
genuine waviness, see above,

It seems that there have been rather few papers exploiting the MS-NS
method to extract information about nuclear scattering. However, the
formal developments by Newton and his successors have been of great
value, not least for showing that there always is a local potential
corresponding to an appropriate radial range for a corresponding
range of partial waves.

An inversion procedure of a similar kind, that due to Lipperheide
and Fiedeldey~\cite{lf,FL84}, LF, was actually the first to extract
information concerning nuclear interactions by means of inversion:
the long range interaction generated by Coulomb excitation of heavy
ions, see~\cite{flf}. With MS-NS or LF $S_l \to V(r)$ inversion,
$V(r)$ is uniquely determined by $S_l$, so prior information must
sometimes be included in the determination of $S_l$. An
application~\cite{allen} to the analysis of scattering data
illustrates this. The resulting potential is compared by Brandan and
Satchler~\cite{brandan} and adjudged less physical than an
alternative described in Section~\ref{discrete}. We emphasize that
this is not a criticism of LF $S_l \to V(r)$ inversion except
insofar as the the inversion procedure~\cite{lf,FL84} requires that
$S_l$ must be represented rather precisely in a specific multi-term
rational function form.

\subsubsection{Semiclassical inversion, WKB methods}\label{WKB}
WKB methods~\cite{kujawski,SLW} are expected to work well at higher
energies. The implementation of these methods is described in the
following papers in which inversions exploiting the WKB
approximations have been carried out, Refs.~\cite{allen1,allen2}. The
WKB inversion procedure has also been exploited in an interesting
study of the effects of systematic errors in the analysis of nuclear
scattering data~\cite{bennett}.

\subsubsection{Iterative-perturbative (IP) inversion}\label{IP}
IP inversion~\cite{mk82,imnp438} exploits the relatively linear
response~\cite{rsmnull,MacKob79} of $S_l$ to changes in $V(r)$ to
construct a procedure based on the iterative correction of a
`starting reference potential', SRP. The SRP in many cases can
be a zero potential. Ref.~\cite{mk82} demonstrated the method
in a calculation of the dynamic polarization potential for the
breakup of $^6$Li. IP inversion was independently developed by
Kukulin~\cite{kp88}. The extension to spin-\hf\ was presented
in Ref.~\cite{imnp467}, the introduction of error analysis in
Ref.~\cite{invprob} and mixed case inversion in
Ref.~\cite{cm91}. More details of IP inversion and its
extensions are given in Section~\ref{IPext}, and applications
will be described in Section~\ref{apps}. The extension of IP
inversion to data-to-potential direct inversion is described in
Section~\ref{direct}.

\subsubsection{Other fixed-$E$ inversion methods}\label{other}
There are various other inversion techniques referred to
in~\cite{kukmac} and~\cite{berlin,honnef,balaton}.
These sources cite many references that are valuable for understanding
the formal issues connected with inversion, but few of the other techniques
seem to have yielded information concerning nuclear interactions.

\subsection{The IP method and its extensions}\label{IPext}
A key feature of IP inversion is
that it is not tied to specific analytic properties of Schr\"odinger's
equation, but simply to the fact that the response of the $S$-matrix to
changes in the potential is approximately linear. This
near linearity leads to properties which give
IP inversion a powerful advantage as a practical tool. These include:\\
(i) It is highly generalizable. Hence, for example, inclusion of
spin-orbit and even tensor forces requiring coupled channel extensions,
are relatively straightforward and do not compromise
the accuracy of the inversion.\\
(ii) Mixed-case and energy-dependent inversion, as
defined above,  are possible.\\
(iii) Useful information can be obtained when the input data are
noisy and incomplete. The iterative procedure can be halted before
S-matrix elements are inverted to a greater precision than is warranted by
their own precision.\\
(iv) IP inversion can be incorporated into a one-step
observable-to-potential inversion algorithm.

The IP approach described here determines a \emph{local\/} potential
corresponding to given $S$-matrices as calculated by Schr\"odinger's
equation; however it has also been applied~\cite{prc40} to determine
a Dirac potential simply by applying the appropriate transformation
to the extracted Schr\"odinger potential.

IP inversion is implemented in the Fortran 90 code Imago. An Imago users's
manual~\cite{imago}, and also Ref.~\cite{kukmac}, give a more general
and detailed account of the formalism than is given below, which presents
the basic idea. An earlier short review of IP inversion and its applications,
including some examples not discussed here, can be found in Ref.~\cite{mcjpg24}.

\subsubsection{Basic IP method}\label{basic}
The IP approach is based on the fact that, in general, the scattering
matrix (or phase shift) responds in a remarkably linear way to changes in
the scattering potential; explicit examples are given in the appendix of
Ref.~\cite{mackob79}. This makes possible a step-wise linearization
procedure in which the potential corresponding to some given $S$-matrix
can be established in a series of iterations starting from a guessed
potential.

The linear response of the $S$-matrix to changes in $V$, which lies at the
heart of the IP method, can be expressed in various equivalent forms. The
change, $\delta S_l$, in the scattering matrix $S_l(k) = \exp{ 2{\rm i} \delta_l(k)}$ for
partial wave $l$ and CM energy $E =\hbar^2k^2/2m$
induced by a small change\footnote{When spin-orbit
interactions are considered, we add relevant labels, e.g. $S_{lj}$.}
in the scattering potential, $V(r)\rightarrow V(r) + \Delta V(r)$, is
\begin{equation}
\delta S_l = {\rm i}\frac{ m}{\hbar^2 k} \int_0^{\infty} \psi^2_l
(r) \Delta V(r) {\rm d}r \label{deles}\end{equation} where
$\psi_l$ is a regular solution of Schr\"odinger's equation
with the asymptotic normalization:
\begin{equation}
\psi_l(k,r)\rightarrow I_l(r) - S_l(k)O(r) \label{asym1}
\end{equation} and $I_l$ and $O_l$ are the conventional incoming
and outgoing Coulomb wave functions. This well known
result follows immediately from the Wronskian
relationships for $I_l$ and $O_l$. These functions can be written in terms of the
regular and irregular Coulomb solutions $F_l = {\rm i}(I_l -O_l)/2$ and
$G_l =(I_l +O_l)/2$. If the Coulomb interaction is absent, $F_l$ and $G_l$
become spherical Bessel $j_l$ and Neumann $h_l$ functions: $ F_l
\rightarrow j_l(kr)$ and $G_l \rightarrow h_l(kr).$

If the target S-matrix to be inverted is $S_l^{\rm tar}$, then the aim is to
determine the potential for which the S-matrix $S_l^{\rm inv} $ renders the sum:
\beq \sigma^2 = \sum_l^{N_l} |S_l^{\rm tar} -S_l^{\rm inv}|^2 \label{sigsq} \eeq
as close to zero as possible, or at least as close as is reasonable. We add that qualifier
since the target S-matrix will have numerical imprecision or errors or superimposed
`noise'; it is a virtue of the IP method that precise inversion can be avoided where
it is inappropriate. Eqn.~\ref{sigsq} omits any labels relating to spin. We shall
also omit labels on components of the potentials (real, imaginary, real and
imaginary spin-orbit etc) which are included in Refs.~\cite{imago} and~\cite{kukmac}.

The minimization is carried out iteratively, starting from
$V_{\rm SRP}$, the \emph{starting reference potential} (SRP).
In favourable cases, the SRP can be a zero potential. An
iteration is carried out as follows: if previous iterations
lead to the current potential $V(r)^{\rm curr}$, the next step
is to find \beq  V(r)^{\rm new} = V(r)^{\rm curr} +
\sum_{i=1}^{N_{\rm b}} \lambda_i v_i(r) \label{new}\eeq where
the $\lambda_i$ are amplitudes to be determined and the
$v_i(r)$ are members of the \emph{inversion basis} of dimension
$N_{\rm b} $. For the inversion basis, the inversion code
Imago~\cite{imago} offers a choice including: displaced
gaussian functions, zeroth order Bessel functions, spline
functions and others. To determine amplitudes $\lambda_i$ such
that the current S-matrix $S_l^{\rm curr}$ may become the
target S-matrix $S_l^{\rm tar} $ (or, at least, closer to
$S_l^{\rm tar} $), we identify $\Delta_l = S_l^{\rm tar} -
S_l^{\rm curr}$ with the change $ \delta S_l$ expressed in
Eqn.~\ref{deles} resulting from the perturbation $\Delta V =
\sum \lambda_i v_i(r)$, to get \beq \Delta_l =  \sum_i^{N_{\rm
b}} \lambda_i \frac{{\rm i} m}{\hbar^2 k} \int_0^{\infty}
\psi^2_l (r) v_i(r) {\rm d}r \equiv \sum_i^{N_{\rm b}} M_{li}
\lambda_i \label{deles1} \eeq where $\psi_l(r)$ is the regular
solution involving the current potential $V(r)^{\rm curr}$. The
next step involves matrix algebra to determine the best set
$\lambda_i$ for the (in general) over-determined system
Eqn.~\ref{deles1}; note that $N_{\rm b} \leq N_l$, the number
of $l$ values included in the inversion. In the original IP
work~\cite{mk82}, and in some subsequent work, we followed
Ref.~\cite{ms} in their use of the standard matrix method.
Using natural matrix notation, Eqn.~\ref{deles1} can be written
$\BfDel = \vect{M} \Bflam$ leading to: \beq \Bflam =
(\vect{M}^\dag \vect{M})^{-1} \vect{M}^\dag \BfDel,
\label{matrix}\eeq with $\vect{M}^\dag $ the hermitian adjoint
of $\vect{M}$. From these values of $\lambda_i$, the new
current potential can be calculated from Eqn.~\ref{new}.
Further iterations can be carried out until a suitable low
level of $\sigma$ is reached.

Before describing the practical implementation, which involves
sequences of iterations rather than a single sequence, we note
that an alternative to the direct inversion of
Eqn.~\ref{matrix} has proven superior: singular value
decomposition, SVD, see, e.g.\ Ref.~\cite{recipes}. SVD makes
convergent iteration possible in cases where the direct matrix
method fails. The first step is to re-write $\vect{M}$ in the
following product form: \beq \vect{M}= \vect{U}
\vect{D}\vect{V}^\dag \label{UDV} \eeq where $\vect{V}$ is
square, $\vect{V}\vect{V}^\dag =1$ and $\vect{U}\vect{U}^\dag
=1$. Matrix $\vect{U}$ will not be square since we are, in
general, dealing with an over-determined system. Matrix
$\vect{D}$ is diagonal with elements $d_j$ for $j =1, \ldots
N_b$. We can then write \beq \Bflam = \vect{V} \vect{D}^{-1}
\vect{U}^\dag \BfDel \label{labSVD} \eeq where $\vect{D}^{-1}$
is diagonal with elements $1/d_j$. In general, $d_j$ vary over
many orders of magnitude. The smallest $d_j$  are the least
accurately determined, and so can be eliminated. The program
Imago does this by setting a tolerance limit, with any elements $d_j$
that are below that limit being set to zero. This limit can be lowered
in successive sequences of iterations as we now describe.

The iterative inversion is not carried out in a single
sequence, but in a series of discrete sequences of iterations.
This allows divergences or oscillatory potentials, following
too large an inversion basis or too small an SVD tolerance, to
be avoided. A sequence of iterations leading to a modest
reduction in $\sigma$ without divergence or spurious
oscillations can be followed by another sequence that has a
basis with a larger $N_b$ or wider radial range, and/or a
smaller SVD tolerance. This will then, typically, converge to a
lower value of $\sigma$. After any sequence of iterations, it
is possible to backtrack if the chosen inversion parameters
lead to divergence or to a potential with oscillatory features
that might be spurious. In practice, depending on the case, one
or several sequences of iterations will lead to a potential
that gives a very close fit to the S-matrix without spurious
oscillations. As implemented in Imago, the fit to both the
S-matrix and the observables can be seen interactively,
on-screen, after each sequence of iterations. Interestingly,
$\sigma$ can often be reduced by an order of magnitude by
further iterations even after a visually nearly perfect fit to
$S_l$ has been achieved over almost the entire $l$-range over
which $1 - |S_l|$ is appreciable. Furthermore, for cases with
many partial waves, a perfect fit to the observables at far
backward angles requires an exceptionally low $\sigma$,
sometimes much lower than required for a good visual fit to
$S_l$.

It is a matter of good practice to test the uniqueness of the
inversion by verifying that the same result is obtained with a
different SRP or inversion basis.

Following a converged inversion procedure, one can be assured
that a potential has indeed been found that reproduces the
input S-matrix to a precision quantified by $\sigma$ and
verified by visual fits to both the S-matrix and the
observables (as mentioned, the latter often being more
sensitive). In addition, the degree of uniqueness can be tested
in the way just described. In the case of input $S_l$ that is
noisy or less well determined, the iterative process can be
stopped at a larger value of $\sigma$ and useful information
extracted, perhaps from a range of alternative solutions. This
can happen in cases of heavy ion scattering at low energies
where there is little useful information in $S_l$ for low
values of $l$ where $|S_l|$ is typically very small.

\subsubsection{Spin-\hf\ inversion}\label{spinhf}
The generalization of IP inversion to the spin-\hf\
case~\cite{npa467}, $S_{lj} \to  V(r) + {\bf l \cdot s}\ V_{\rm
SO}(r)$, is straightforward and its implementation in the code
Imago is described if Ref.~\cite{imago}.
For spin-\hf\ scattering the potential becomes,
\begin{equation}
V_{\rm cen}(r) + iW_{\rm cen}(r) +
2{\bf l \mbox{\boldmath $\cdot$}s}\,(V_{\rm
so}(r) +i W_{\rm so}(r)).\label{v-def}
\end{equation} With suitable indexing, it is straightforward
to expand the matrix system to accommodate the new terms. An
independent inversion basis for the spin-orbit term can be defined
and may differ from that for the central term in two respects: (i) it does not extend quite
to the origin, since only the $l=0$ radial wave function is
non-zero there and ${\bf l \cdot s}$ is zero for $l=0$, and,
(ii) it need not extend to such a great radius, since, in fact,
spin-orbit terms tend to be small at a radius where central
terms are still finite.

\subsubsection{Mixed case and energy-dependent inversion}\label{mixed}
Mixed case inversion determines the potential that reproduces $S_{ls}(E_i)$
for a few discrete energies $E_i$. This is particularly useful at low energies
and with light target nuclei when only a few partial waves are active  --- too few to
define the potential for a single energy. It is quite straightforward to
expand the matrix system to include multiple energies and multiple sets
of target $S_{lj}$. Because the nuclear optical potential is intrinsically
energy dependent, mixed case inversion is useful only over relatively
narrow ranges of energy, unlike energy dependent inversion.

Energy dependent IP inversion allows the parameters of postulated energy dependent
functions to be optimized in an expansion of the matrix system described above.
Ref.~\cite{prc54} describes the formulation as originally applied and as is
implemented in Imago. A significant feature is that different parameterized forms
must be used for the real and imaginary components. This is obviously important
in the case discussed in Ref.~\cite{prc54} where the energy range involved
crossed the inelastic threshold for p + $^4$He scattering.

\subsubsection{Parity dependence and identical bosons}\label{parity}
Parity-dependent inversion determines, in effect, separate interactions
for the even-parity and odd-parity partial waves.  Parity-dependent terms
can also be straightforwardly generated by including
a Marjorana inversion basis $v_i^{\rm M}(r)$ so that any term x, where x could refer to
real central, imaginary central, real spin-orbit or imaginary spin-orbit,
can have an added Majorana term of the form
$(-1)^l \sum_i \lambda_i^{\rm M} v_i^{\rm M}(r) $.

Inversion for the scattering of identical bosons involves inversion for which
$S_l$ exists only for even $l$. The only issue for IP inversion is whether
there are enough partial waves to define the potential with sufficient accuracy.
Section~\ref{discrete} describes several cases of inversion involving
identical bosons at fairly high energies for which there were plenty of partial
waves for satisfactory inversion. For low energy
identical bosons involving $S_l$ calculated from theory, it might be possible
to interpolate to odd $l$.

\subsubsection{Including bound state energies}\label{bound}
It is possible to include the energies of bound states as input to the inversion
process. This will be useful at low energies where, even with an S-matrix
for multiple energies in energy dependent inversion, there is a paucity of information
with which to define the potential. The method was introduced by Cooper and fully
described in~\cite{prc50}. In that reference, a fit to the energy of an $l=1$
bound state in $^7$Be was included in an inversion of low energy $S_{lj}$ for $^3$He
scattering on $^4$He, leading to a parity-dependent potential.

\subsubsection{Spin-1 inversion yielding tensor
interaction}\label{spin1} The extension of the IP algorithm to spin-1
inversion is not as straightforward as the extensions described above.
This follows from the fact that the scattering of spin-1 projectiles from
a spin-zero target requires, in general, a coupled channel calculation.
For example for total angular momentum $j =1$ and positive parity, the $l=0$ and
$l=2$ channels may be coupled by a tensor force interaction. The general S-matrix
for spin-1 projectiles scattering from a spin-zero target may be written $S^{j}_{ll^{\prime}}$
where $l= j-1, j\mbox{ and } j+1$. Channels with $l= j-1 \mbox{ and } j+1$ will, in general,
be coupled. The possible
tensor interactions were classified by Satchler~\cite{np21, satchler} and labeled $T_{\rm R}$,
$T_{\rm P}$ and $T_{\rm L}$. The third of these is believed to be small,
and the $T_{\rm P}$ interaction appears to be hard to distinguish phenomenologically
from the first, $T_{\rm R}$, interaction. The $T_{\rm P}$ interaction might be
important, but involves gradient operators for which an inversion procedure
has not been devised. We therefore assume that the inter-nuclear interaction
may contain a tensor force component of the form:
\beq T_{\rm R}V_{\rm R}(r) \equiv (({\bf s \cdot \hat{r}})^2 -2/3)V_{\rm R}(r).
\label{tr}\eeq
To invert an S-matrix of the form $S^{j}_{ll^{\prime}}$ to determine a potential including
a $T_{\rm R}$ interaction, requires coupled channel inversion in which a non-diagonal
S-matrix yields a non-diagonal potential. A specification of the formalism together
with an account of tests of its performance can be found in Ref.~\cite{npa677}. This
reference also presents the results of inverting $S^{j}_{ll^{\prime}}$ generated with
a $T_{\rm P}$ interaction leading to a $T_{\rm R}$ interaction. The tests in Ref.~\cite{npa677}
were for a single energy case with no parity dependence, but did include an extension to
direct data-to-potential inversion, see Section~\ref{direct}.

Spin-1 inversion leading to a $T_{\rm R}$ interaction is straightforwardly extended
to include parity dependence, Section~\ref{parity}  in all components as well as
energy dependence, Section~\ref{mixed}. An application of this very general inversion
to deuteron scattering from $^4$He at low energies is presented in Ref.~\cite{npa723}.

\subsubsection{Further extensions?}\label{further}
Spin-1 inversion, just discussed, is a restricted form of coupled
channel inversion in which a non-diagonal S-matrix is inverted to
give a non-diagonal potential of a very specific form. Just how far
the IP concept can be pushed to provide a more general form of
coupled channel inversion is a challenge for the future. The
difficulty of providing experimental data of sufficient breadth and
precision might restrict the application of coupled channel inversion
to non-diagonal S-matrices that have been calculated from theory.

\section{Applications of inversion in nuclear scattering}\label{apps}
There are a variety of good reasons for determining local potentials
that reproduce given sets of S-matrices or phase shifts, and some of
these will emerge from the account given here of the various
applications of inversion. We cannot give an
exhaustive list of possible applications, and readers may well be inspired to
find new ones.

\subsection{Nucleon-nucleon and similar interactions}\label{NN}
Unlike nucleon-nucleus potentials, nucleon-nucleon interactions are
not generally considered to be explicitly energy-dependent and are
therefore a candidate for the application of fixed-$l$
phase-shift-to-potential inversion. This would exploit the
comprehensive phase-shift analyses covering a wide energy range. In
fact `fixed-$l$' is a misleading term since the tensor interaction
mixes $l$ channels for given conserved total angular momentum and
parity. The $l$-dependence of the potential and the small number of
active partial waves effectively rule out fixed-$E$ inversion.

The challenge of applying Gel'fand-Levitan-Marchenko methods,
generalized to allow for coupled channels inversion to determine the
non-diagonal tensor interaction, was successfully taken up by von
Geramb and collaborators, see their articles~\cite{honnef}.

\subsection{`Two-step' nuclear elastic scattering phenomenology.}\label{twostep}
It is often desirable to fit elastic scattering observables with
potential models that have as few as possible \emph{a priori}
assumptions (or prejudices) concerning their nature. To achieve
this, a number of `model independent' fitting algorithms to
determine optical model (OM) potentials, typically based on sums of
gaussian or spline functions, have been developed. These often allow
point-by-point uncertainties to be assigned to the potential.

The possibility of inversion affords an alternative approach and
arguments in support of this approach have been
made~\cite{rsmnull,prc45,npa552,npa576,npa582}. The idea is to first
determine $S_l$ or $S_{lj}$ by fitting the elastic scattering
observables (angular distributions (ADs) analyzing powers (APs)
etc). These $S_l$ or $S_{lj}$ can then be inverted in a subsequent
step. We refer to this overall procedure as `two-step nuclear
(elastic scattering) phenomenology'. We here distinguish two classes
of two-step phenomenology: (1) determination of the S-matrix at one
or a few discrete energies, and, (2) the fitting (usually for
few-nucleon target nuclei) of  functional forms $S_l(E)$ or $S_{lj}(E)$,
over a fairly continuous range of energies, by means of an R-matrix or effective
range procedure. We discuss these two cases separately, and show
that they unambiguously reveal the parity dependence of many
light-ion nucleus-nucleus interactions.

\subsubsection{Discrete-energy OM fitting by inversion}\label{discrete}
When elastic scattering data, particularly data of typical precision
and angular range, are fitted by searching on S-matrix elements,
various profoundly different solutions may be found. Each S-matrix
solution will lead, by inversion, to a different potential. In
fact, the experience of carrying out such searches is very revealing
about the under-determination of the potential by the data. Even
so-called `model-independent' OM fits often implicitly embody prior
information that appears to ameliorate, but not completely remove,
this problem. S-matrix searches at single energies, therefore,
should be constrained with prior information; this is possible. We
note a study of the effects of systematic errors in the analysis of
nuclear scattering data~\cite{bennett}.

There are various ways of incorporating prior information. One can
start with an analytic form for $S_l$ (e.g. McIntyre, Wang and
Becker~\cite{mwb} (MWB)) and search on the
parameters~\cite{prc45,npa552}, or one can start from an MWB or
similar analytic form $S(l)$ and search on an additive component
$\Delta S(l)$ in a way that preserves unitarity. Generally, it is
important to regulate or constrain the search in appropriate ways.
Useful, physically motivated, starting functions $S(l)$ for the
S-matrix search are the S-matrices calculated by a Glauber model or
by an existing phenomenological fit, see~\cite{npa576}.

In Ref.~\cite{prc45}, the elastic scattering of $^{16}$O on $^{12}$C
at 608 MeV was analyzed by first determining $S_l$ by fitting the
elastic scattering data by means of an additive correction to the
$S_l$ of the McIntyre, Wang and Becker (MWB)~\cite{mwb}
parameterized form. The correction was a searchable spline function
of $l$. The angular distribution was first approximately fitted with
a five parameter MWB form $S_l$; the subsequent fitted spline
function addition led to a threefold reduction in $\chi^2/N$. The
resulting corrected MWB-derived potential revealed a very different
degree of surface transparency compared to the uncorrected MWB
potential. At this energy the IP inversion is precise and stable,
leading to effectively identical potentials independently of whether
the iterative inversion of $S_l$ started from a \emph{zero} potential,
$V(r)=0$, or a Woods-Saxon potential in the neighbourhood of the expected
result.

Ref.~\cite{npa552} discusses in depth the application of two-step
phenomenology in an account of its application to $^{12}$C +
$^{12}$C elastic scattering at 9 energies from 140 to 2400 MeV. This
paper, in effect, presents a critique of standard OM phenomenology,
with a discussion of the advantages (including computational
efficiency) of two-step phenomenology and the means of implementing
it. The strategy for avoiding spurious solutions was discussed,
including the application of continuity-with-energy to the
solutions, which, in this case, revealed apparent serious
shortcomings of the data at one specific energy. This study also
revealed weaknesses in the conventional Woods-Saxon phenomenology
across the whole energy range. At these energies, there is no
problem in inverting $S_l$ for just even values of $l$, as arises with the
scattering of identical bosons.

An analysis of $^{16}$O + $^{16}$O elastic scattering was
carried out~\cite{npa576}, fitting wide angular range and
precise data at a single energy, 350 MeV. Alternative solutions
for $S_l$ starting from the S-matrices calculated by a Glauber
model or an existing phenomenological fit, led to very precise
fits for $\sigma(\theta)/\sigma_{\rm Ruth}(\theta)$ spanning
nearly 5 orders of magnitude, leading to very similar
potentials for $r \ge 4$ fm. There is little sensitivity at
radii less than that, contradicting claims for repulsive effects at a
small radius. Brandan and Satchler~\cite{brandan}, their
Section 9.2, evaluate the resulting potentials~\cite{npa576} in
comparison with potentials obtained using alternative
procedures. We note in passing that the simplified Glauber
model gave $|S_l|$, but not $\arg{S_l}$, in close agreement
with values from the final fitted $S_l$.

In Ref.~\cite{npa582} two-step inversion is applied to the elastic
scattering of $^{11}$Li from $^{28}$Si (at 319 MeV) and from
$^{12}$C (at 637 MeV). This work revealed the profound ambiguities
in $S_l$, and thus in $V(r)$, that occur when fitting data of
limited range and precision. These ambiguities are more extreme than
those that are found with conventional OM fitting. Good fits to
the data were easily (too easily?) obtained in which the large-$r$
tail was surprisingly extended, although this possibly results from
contamination of the forward angle elastic AD data with inelastic
scattering. Realizing the full potential of the two-step method for
analyzing the elastic scattering of halo nuclei awaits the advent
of sufficiently high quality data.

Inversion can, of course, be applied to an independently fitted
published S-matrix. In Ref.~\cite{prc40} IP inversion of $S_{lj}$
for p + $^4$He elastic scattering was carried out for $S_{lj}$ that
had been fitted to angular distribution data at a single energy,
64.9 MeV. The resulting potential was not of Woods-Saxon-like form,
but exhibited long-wavelength oscillations. The reason for these
became evident later with subsequent multi-energy inversions, see
Section~\ref{rmat}. Ref.~\cite{prc40} also presented real and
imaginary, scalar and vector, Dirac equivalent to the Schr\"odinger
potentials. This demonstrated that, with certain limitations, Dirac
equation $S \to V$ inversion is possible by way of Schr\"odinger
equation equivalence.

\subsubsection{Inverting S-matrices from R-matrix and effective range fits}\label{rmat}
Mixed case and energy-dependent inversion make possible an
alternative form of two-step inversion. The first step now takes the
form of an R-matrix or effective range fit to elastic scattering
data over a possibly quite wide energy range, which might include
shape resonances. The result is an analytic form of S-matrix in
which the (typically) small number of active partial waves is
compensated for by the fact that $S_{lj}(E)$ exists for a
substantial range of energy $E$. Satisfactory inversion becomes
possible even in cases where the energy range is much less than is
required for fixed-$l$ inversion and, also, the number of partial
waves is much fewer than what is required for fixed-$E$ inversion.
Inversion of this kind becomes particularly interesting when the
results can be compared with the potentials derived from the
inversion of $S_{lj}(E)$ from RGM or similar theories for the same
few-nucleon systems; we discuss the specific case of nucleon
scattering from $^4$He in Section~\ref{rgm}.

In Section~\ref{discrete} it was mentioned that single-energy
two-step inversion for p + $^4$He led to wavy potentials.
Multi-energy inversion provides an explanation and points to a
significant property of nucleus-nucleus interactions between
few-nucleon nuclei. Ref.~\cite{cm91} applied mixed case inversion to
$S_{lj}$ for p + $^4$He that had been fitted, using R-matrix and
effective range expansions, to experimental data at several discrete
energies. As a result, the following alternative emerged:
\emph{either} (i) the potential is wavy (as it was for 64.9 MeV),
\emph{or}, (ii) there is a smooth but parity-dependent potential.
The parity dependence is such that the odd-parity potential has a
substantially greater range and volume integral than the even-parity
potential. We shall see in Section~\ref{rgm} that exactly this form
of parity dependence emerges from the inversion $S_{lj}$ derived
from theories that include exchange processes (excluding knock-on,
Fock term, exchange). Ref.~\cite{cm91} found that the p + $^4$He and
n + $^4$He \emph{nuclear} interactions were essentially identical in
the odd-parity channels (as required by charge symmetry) but
differed somewhat for $r \le 2$ for the even-parity channels.

In Ref.~\cite{cm91}, only energies below the inelastic threshold were
involved so the potentials were real, unlike those of
Ref.~\cite{prc40} or those found in Ref.~\cite{prc54} in which
energy dependent IP inversion was introduced. Potentials for p +
$^4$He scattering were found by inverting $S_{lj}(E)$ from various
R-matrix and effective range fits to experimental data from zero
energy to about 65 MeV. The resulting potentials fitted the shape
resonances at energies below the inelastic threshold, and became complex
above the threshold, reproducing the data reasonably well up to 65 MeV.
The opening of a relatively small number of specific channels at various
energies above the threshold revealed the limits of potential models
in which the potential varies smoothly with energy. Such a model
evidently requires either the `many open channel' condition of the
standard optical model, or, the `zero open channel' situation that
holds below threshold; neither is true between the threshold and 65
MeV for p + $^4$He. Refs.~\cite{cm91,prc54} together demonstrated,
on an empirical basis, that the real (and imaginary above threshold)
central as well as spin-orbit components of the nucleon-$^4$He
potential are parity dependent; they also established the
practicality of energy-dependent inversion. The key finding: over
the whole energy range considered, the parity dependence of the real
central part is such that the odd-parity component has both a longer
range and a greater volume integral than the even-parity term. Thus,
potentials that have a factor $(1 + \alpha (-1)^l)$ multiplying a single
radial form (as have been applied in optical model fits) are too
restrictive.

Parity dependence extends beyond 5-nucleon systems: IP inversion of
$S_l$ that had been fitted~\cite{plaga} to $^4$He + $^{12}$C elastic
scattering data yielded~\cite{npa517} a strongly parity-dependent
potential that reproduced the scattering data very well, including
the shape resonances. The potential differed in the surface region
from previous phenomenological potentials that had been found in a
conventional way; this is of possible astrophysical significance.

The $^3$He + $^4$He interaction, which is also of astrophysical
importance, was shown to be strongly parity-dependent by Cooper in
Ref.~\cite{prc50} in which both empirical and theoretical $S_{lj}$
were inverted. The bound state energy was also included as input to
an IP inversion for the first time, contributing to the
determination of the interaction.

\subsection{Data-to-potential direct inversion}\label{direct}
It is possible to combine the determination and the IP inversion of the
S-matrix into one algorithm, see
Refs~\cite{yadf,npa618,npa645,npa677,npa723}. Scattering
data that has been measured for multiple energies can be included,
in this way implementing energy-dependent, direct data-to-potential
inversion.

In Ref.~\cite{npa618}, for protons scattering from $^{16}$O at 7
energies from 27.3 MeV to 46.1 MeV, very precise wide angular range
AD and AP data were fitted using energy-dependent direct
data-to-potential inversion. Parity-dependent real and imaginary
central potentials and complex Wigner spin-orbit potentials were
determined. The even-parity and odd-parity central potentials were
smooth and behaved in a regular way with energy. The odd-parity
(Majorana) terms were very like those found from the inversion,
described elsewhere, of $S_{lj}$ derived from RGM calculations
of protons scattering from $^{16}$O. Fits of equal quality lacking
parity dependence would certainly require wavy potentials.
This work by Cooper might reasonably be described as
state-of-the-art nucleon scattering phenomenology for a single
nucleon-nucleus pair; it conclusively establishes the parity
dependence of the interaction between a nucleon and $^{16}$O. As a
result, we conclude that the omission of parity dependence from
tests of folding model theories, as applied to nuclei as light as
$^{16}$O, will lead to misleading results.

In Ref.~\cite{npa645} a parity-dependent potential, including spin-orbit
terms, that gave a fair simultaneous fit to ADs and APs for $^6$Li +
$^4$He scattering at 19.6, 27.7 and 37.5 MeV, was found. The
Majorana terms were essential for the fit in this few-nucleon
system. This appears to be something one must presume to be required
in all few-nucleon system inversions, see Section~\ref{rgm}.

In Ref.~\cite{npa677}, which introduced coupled-channel IP inversion for
the scattering of spin-1 projectiles leading to a $T_{\rm R}$ tensor
interaction, data-to-potential inversion was carried out fitting
multiple energy data including angular distributions, vector
analyzing powers and the three tensor analyzing powers for deuterons
scattering from $^4$He. The energy range was from 8 to 13 MeV for 6
discrete energies; 4000 data were fitted with a potential that
included parity-dependent central and tensor terms. A subsequent
study~\cite{npa723} fitted a wider range of energies, including the
three D-state resonances. Strong, complex parity-dependent $T_{\rm
R}$ tensor interactions were revealed.

Direct data-to-$V$ inversion of data covering a
substantial range of energies, leading to a $T_{\rm R}$ tensor interaction,
with all components parity-dependent  (where required by the data) and
energy dependent, represents the most complete implementation of
the IP inversion procedure.

\subsection{Determination of potentials from theory}\label{theory}
The inversion of S-matrices calculated from a theory is not subject to
the problems that may arise, e.g.\ from noise and experimental
uncertainties, when the S-matrix determined from scattering data is
inverted. Applications of the inversion of calculated S-matrices
include: (i) determining in a
natural and efficient way, the local potential that is S-matrix equivalent
to non-local or $l$-dependent potentials; (ii) deriving potentials
that represent the scattering for those theories
(see e.g.\ Section~\ref{rgm}  and Section~\ref{kmt}) that calculate
scattering directly without the intermediary of potentials; (iii)
providing arguably the best method of calculating the dynamic
polarization potential (DPP) due to the coupling of inelastic or
reaction channels to the elastic channel.

\subsubsection{Dynamic polarization potentials (DPP) by CC-plus-inversion}\label{dpp}
The contribution of inelastic processes to nucleus-nucleus
interactions is represented by the dynamic polarization
potential (DPP)~\cite{brandan,satchler} the non-local and
$l$-dependent form of which (for inelastic scattering, at
least) was derived by Feshbach~\cite{feshbach}. This has been
calculated with various approximations, see for example
Refs.~\cite{CS77,rawit87}, and leads to a highly non-local and
$l$-dependent expression. Moreover, the inclusion of coupling
to transfer channels with full finite range coupling and the
inclusion of non-orthogonality terms has never been achieved in
such calculations. Finally, the results are not easily related
to phenomenology since it is necessary to establish local and
$l$-independent equivalent potentials from such calculations and
this requires the calculation of $S_{lj}$ from non-local and
$l$-dependent potentials.

There is an alternative procedure for calculating DPPs that can
handle reaction channels, coupling of all orders and (in principle)
exchange processes: `coupled channel plus inversion'. In
this method, a potential is first found by inverting the elastic channel S-matrix
from a coupled channel calculation. When the bare 
potential of the CC calculation is subtracted from this inverted
potential, the resulting  difference potential is a local
and $l$-independent representation of the DPP arising from the channel 
coupling. A full discussion of this procedure, its advantages and limitations, 
and a comparison with other methods, is given in Ref.~\cite{prc81}.
The method used in earlier attempts to extract the contribution
of channel coupling to the optical potential (see Ref.~\cite{npa164}
and references cited there) was to refit the CC angular distributions,
but this is subject to the many limitations of optical model fitting.

Many processes that contribute to the DPP can be studied with
CC-plus-inversion: coupling to inelastic channels, reaction
channels, particle or cluster exchange and projectile breakup.
IP inversion was introduced in a study~\cite{mk82} of the
contribution of projectile breakup to the $^6$Li-nucleus
interaction. Many other studies of the DPP due to breakup of
$^6$Li have been made; Ref.~\cite{impl86} described generic
properties of the DPP that were common to the breakup of
deuterons and $^6$Li. A recent paper on $^6$Li breakup,
Ref.~\cite{pangm} which has many references, compared DPPs
derived using $S$-matrix inversion and those from an
$l$-weighted TELP, see Section~\ref{alt}, and found marked
differences. The DPP due to the breakup of the halo nucleus
$^6$He has also been studied~\cite{prc70,prc71,prc79}: both the
real and imaginary parts have remarkably long tails
attributable to Coulomb breakup. The tail on the imaginary part
is absorptive for $r\ge 13$ fm but emissive for smaller $r$;
the real part is attractive at large radii, but with a sharp
change to repulsion at about 15 fm. Potentials are presented
out to 60 fm. In Ref.~\cite{npa834} DPPs are compared for the
breakup of $^6$Li, $^7$Be and $^8$B scattering from $^{58}$Ni.
Breakup in which the Coulomb interaction plays a much smaller
role was evaluated for protons scattering from $^6$He in
Ref.~\cite{prc67}.

A stimulus to the development of IP inversion was the
discovery~\cite{npa209230,plb44,plb62} that the coupling to deuteron
channels appeared to have a large effect on proton scattering. This
raised the question of what contribution this coupling makes to the
nucleon OMP. An early application~\cite{plb178}
of spin-\hf~\cite{npa467} IP inversion presented the effects of finite
range p$\Leftrightarrow $d coupling  on the real and imaginary, central
and spin-orbit terms of the p + $^{40}$Ca OMP at 30.3 MeV. The effect
of finite-range coupling (not included in the earlier work) on the DPP
was shown and the individual and total contributions of lumped
$\frac{5}{2}$+, \hf+ and $\frac{3}{2}$+ pickup states was presented.
Later studies of the contribution to the proton OMP of
p$\Leftrightarrow $d coupling, including non-orthogonality terms
previously omitted, are referenced in Ref.~\cite{prc81}; these include
cases of proton scattering from halo nuclei, most recently~\cite{prc83a,prc83b}.

In Ref.~\cite{mcinp472} it was found that the coupling mass-three
channels had a major effect of deuteron elastic scattering; inversion
revealed large repulsive DPPs. Subsequent development of CRC codes
permit the inclusion of non-orthogonality corrections and finite range
interactions, not included in Ref.~\cite{mcinp472}, and, at the same
time, spin-1 inversion leading to the $T_{\rm R}$ had been developed.
These advances were exploited in Ref.~\cite{prc77} which determined
the real and imaginary, central, spin-orbit and tensor DPPs generated
by coupling to mass-3 pickup channels for 52 MeV deuterons scattering from
$^{40}$Ca. The volume integral of the real, central DPP is much smaller than
before~\cite{mcinp472}, but the magnitude point-by-point is not small,
reflecting the wavy character of both real and imaginary components.
This is indicative of $l$-dependence, Sect.~\ref{ldep}, and could not
have been picked up by refitting the elastic scattering angular distributions
from CC calculations, as in Ref.~\cite{npa164}.

{\bf J-weighted inversion.} Inversion has not been developed for spin $>1$
and Ref.\cite{prc77} introduced and tested a means of achieving
inversion leading to a meaningful central potential for projectiles
with large spin. The idea is to define a J-weighted S-matrix:
\begin{equation} \bar{S}_l = \frac{\sum_J(2J+1)
S^J_{ll}}{\sum_J(2J+1) \label{sbar}} \end{equation}
that could be inverted in the usual way. It was found that for the spin-1
case studied in Ref.\cite{prc77}, the imaginary part of the J-weighted DPP
was close to the imaginary part of the central DPP from the complete inversion.
The real part was qualitatively reproduced. The J-weighted procedure was later
applied to the breakup of $^6$Li, $^7$Be and $^8$B scattering from $^{58}$Ni
mentioned above, Ref.~\cite{npa834}.

\subsubsection{RGM, GCM and other few-body cases; exchange contributions}\label{rgm}
Inversion can play a particular role in supporting our
understanding of the scattering of few-nucleon systems. Even a
scattering system as simple as a nucleon plus $^4$He becomes very
complicated if all reaction channels, realistic NN interactions
and a full account of exchange processes are to be included. At
least until recently, the standard methods of calculating
scattering observables would be the application of resonating
group methods (RGM) or generator coordinate methods
(GCM)~\cite{wildertang,slyv}. Even for systems of 4 or 5
nucleons, greatly simplified nucleon-nucleon interactions,
generally omitting tensor terms, are employed. How are such
theories to be tested in view of the inevitably approximate
fits to data? Inversion provides a partial solution. In
Section~\ref{rmat} we described the inversion of S-Matrices
that had been fitted to experimental observables and noted that
definite qualitative features of the potentials emerged, e.g.\
parity-dependent potentials with specific differences between
the strength and range of the even-parity and odd-parity terms.
Such features are a consequence of the various exchange terms
that are precisely those aspects of the scattering process that
RGM and GCM calculations treat correctly. The inversions to be described
support very well the general conclusions of RGM and GCM
calculations. In particular, it becomes possible to study the
contributions of the many different exchange processes, apart
from knock-on exchange (c.f. Section~\ref{ldep}), that are
included in RGM-GCM calculations, including those that are
responsible for parity dependence. LeMere
\etal~\cite{lemere,lemere1} discussed in qualitative terms the
nature of the contributions specific exchange terms would make.
Moreover, Baye~\cite{baye} has made predictions concerning the
way the effect of exchange processes, including those leading
to parity dependence, depends upon the masses of the
interacting nuclei. It is important to check such things,
particularly since, as we have already seen,  parity dependence
is substantial for neutrons or protons scattering on $^4$He
and apparently important for p + $^{16}$O. Is it still
important for n + $^{40}$Ca? Baye's work would suggest not, but
we need to know since its presence or absence would make a
difference in precise checks of folding model theory for n +
$^{40}$Ca.

The p + $^4$He case is of particular interest for two reasons:
it is less intractable than most, and clear parity
dependence has been found from the analysis of experimental data,
as reported in Section~\ref{rmat}.
RGM calculations of $S_{lj}$ for p + $^4$He scattering, in some of
which d + $^3$He configurations were included, were inverted
using mixed case inversion. The calculations were below
threshold so the potentials were real. The inversion
confirmed the key result from the inversion of empirical R-matrix
phase shifts, that the potential shows strong parity dependence
such that the odd-parity potential is of considerably greater
range and volume integral than the even-parity term.
More elaborate p + $^4$He RGM calculations extending
above the inelastic threshold  with coupling to the d + $^3$He
channels were presented in Ref.~\cite{npa626}. The contribution of
the breakup of the deuteron in the d + $^3$He channels was studied.
It was found that the reaction channels increased the Majorana term
in the proton potential. In this work, the contribution of the
p + $^4$He channel to the d + $^3$He interaction was determined
for $S= \hf$ and $S = \frac{3}{2}$ channel spins.

A challenge for energy-dependent IP inversion was the
inversion~\cite{npa742} of  $S_{lj}$ from 0 to 25 MeV for
p + $^6$He, calculated~\cite{pdes} using RGM. No reaction channels
were included, but parity dependence was allowed for in the inversion.
Spin-orbit terms were also included. Two specific questions were posed: (i)
is the energy dependence of the real, central Wigner term consistent
with the energy dependence of the global optical potential? and,
(ii) is there a Majorana term that is less than that for p + $^4$He
in a way that is consistent with the predictions of Baye~\cite{baye}?
Inversion yielded a single parity-dependent potential with an energy-dependent form
that fitted $S_{lj}(E)$ over essentially the whole energy range, and
verified both points (i) and (ii). Point (i) is interesting since
it shows that the global OM energy dependence extends down to mass 6. Since
there were no reaction channels in the RGM, it suggests that, as widely
assumed but sometimes doubted, the bulk of the energy dependence of the
nucleon-nucleus interaction is, indeed, a product of knock-on exchange.

In Ref.~\cite{npa592}, RGM S-matrices for the following cases
were inverted and the potentials evaluated: p + $\alpha$, n +
$\alpha$, p + $^3$He, n + $^3$H, n + $^6$Li, n + $^{16}$O and n
+ $^{40}$Ca. For the cases where the channel spin exceeded \hf,
potentials (without spin-orbit components) were determined
separately for each value of the channel spin. The Majorana
terms were generally large, and in cases where there were two
values of channel spin the Majorana term was quite different
for each value, usually in a way for which plausible physical
reasons exist, in line with what had been
suggested~\cite{lemere,lemere1}. This work verified the parity
dependence of the p + $^{16}$O interaction, and also found very
little parity dependence for n + $^{40}$Ca, as long as the
$l=0$ partial wave was excluded. Apart from this last proviso,
this is in close accord with the predictions of
Baye~\cite{baye}.

The contributions, together and separately, of specific
exchange terms for $^4$He + $^{16}$O, $^3$He + $^4$He and $^3$H
+ $^4$He, were studied by inversion in Ref.~\cite{npa589}. Of
the various conclusions that were drawn, we mention just one:
when a parity-\emph{in}dependent purely phenomenological imaginary
potential is included in the RGM calculation, in order to
enable a more reasonable comparison with experiment, the
imaginary part of the inverted potential is parity dependent.
This might be due to the fact that the real potentials for each
parity had differing degrees of non-locality and hence
differing Perey effects. Perey effects are discussed in
Section~\ref{ldep} below where the effect of non-locality on
the inverted imaginary potential is noted.

S-matrices, $S_l$, from RGM calculations~\cite{wada} for
$^{16}$O + $^{16}$O elastic scattering, for seven energies from
30 MeV to 500 MeV, were inverted~\cite{npa562} using IP
inversion. Since the direct (non-exchange) potential was fixed,
this provided an energy-dependent local equivalent of the
interaction generated by the exchange term. The
authors~\cite{wada} had determined $V(r)$ from the the RGM
$S_l$ using $l$-weighted WKB inversion, so the IP inversions
constituted a test of $l$-weighted WKB inversion; this appeared
to work quite well for energies of 150 to 300 MeV, but was not
accurate at the lower energies studied, 30 - 59 MeV. At 150 MeV
and below, the exchange terms are quite strongly repulsive at
the nuclear surface but attractive at the nuclear centre. These
effects are much smaller at the higher energies, there being no
surface repulsion at 500 MeV.

\subsubsection{Potential representation of KMT and comparable theories}\label{kmt}
Other theoretical formulations of nuclear scattering also produce an
S-matrix directly, without the intermediate stage of generating
a potential. Such methods do not as yet give perfect fits to
observables, so it is of interest to compare the local
potential that would reproduce the $S_{lj}$ of the formalism
with the well-established phenomenological OMP. An example of
such a theory is that due to Kerman, McManus and Thaler~\cite{kmt},
KMT. More recent developments of KMT theory have included
multiple scattering terms and Pauli blocking effects. In
Ref.~\cite{prc49} first and second order KMT potentials for
nucleon-$^{16}$O scattering are calculated at 100 and 200 MeV
by applying IP inversion to the corresponding KMT S-matrix. The
volume integrals and rms radii of the four components of the
inverted potentials are given, facilitating a comparison with
established global phenomenology as well as with the results of
alternative theories based on local density approximation
nuclear matter theory. Ref.~\cite{prc49} gives references to
earlier KMT calculations by the first three authors and others.

\subsubsection{Inverting $S_{lj}$ from non-local and explicitly
$l$-dependent potentials}\label{ldep} {\bf Non-locality:} IP inversion, whether
energy dependent or not, is a convenient means of determining
the local equivalent of a non-local potential. The non-locality
in the nucleon-nucleus interaction that is due to inelastic
processes is not well established, but the non-locality due to
knock-on exchange (Fock term) is well known, and is the major
source of the energy dependence of the local OMP. The energy
independent Perey-Buck~\cite{pb} non-local potential, which
fits nucleon elastic scattering over a wide energy range, is
thought to represent the exchange non-locality in a simple
parameterized way. If the $S_{lj}$ for a non-local potential
can be calculated (this is straightforward), then IP inversion
immediately yields the local-equivalent potential. If
$S_{lj}(E)$ is calculated over a wide range of energies, then
energy-dependent IP inversion immediately yields the energy
dependence of the local equivalent potential. This has been
done~\cite{jpg23} for the Perey-Buck potential, leading to a
calculated energy dependence that well matches the energy
dependence of the empirical OMP. IP inversion was also applied to
$S_{lj}$ that had been calculated from a potential in which the real part was
of Perey-Buck non-local form but in which the imaginary part
was local. The resulting local potential had an imaginary term
that was reduced compared to that included with the non-local
real potential; the reduction factor was just the
Perey~\cite{perey} reduction factor of the wave function within
the nucleus that is due to the non-locality of the real
potential.

The  S-matrix for a microscopic non-local calculation of neutron-$^{16}$O
scattering was inverted in Ref.~\cite{prc49rawit}. It was found that the
damping of the wave function within the nucleus, following from an exact
microscopic treatment of exchange, matched very closely the damping
associated with the phenomenological Perey-Buck potential: the original
Perey effect~\cite{perey}. This leaves surprisingly little room for
damping from reaction and inelastic processes suggesting, maybe
in line with Austern's picture~\cite{austern}, that the non-locality arising
from channel coupling redistributes flux, but this occurs without a
global reduction in the magnitude of the wave function.

{\bf $l$-dependence:} IP inversion also provides a means of finding
the $l$-independent equivalent to an explicitly $l$-dependent potential; this must
exist. Parity dependence is not the only form of $l$-dependence
that has been proposed, on various grounds, as a property of
phenomenological OMPs (the Feshbach theoretical potential is
$l$-dependent and also see Refs.~\cite{pl68b,km79,km81}). As we
have emphasized, the S-matrix $S_l$ that is calculated from an
$l$-dependent potential can be inverted to yield an
$l$-independent equivalent. There is a motivation for doing so:
if model-independent OM phenomenology produced a local potential
of a form that was recognizably equivalent to an $l$-dependent
potential, that might be regarded as evidence that there is
`really' $l$-dependence of that form. As an example, in
Ref~\cite{npa542}, $S_l$ from $l$-dependent potentials that had
been fitted to $^{16}$O + $^{16}$O scattering for energies from
30 MeV to 150 MeV were inverted. The real and imaginary
$l$-dependencies were of quite different forms according to
quite different physical motivations. The inverted potentials
varied with energy in a systematic way. However, the imaginary
part in particular was quite unlike any found with standard
optical model phenomenology, casting some doubt on the
particular $l$-dependence of the imaginary potential introduced
by Chatwin \etal~\cite{chatwin}.

In Ref.~\cite{prc76}, local and $l$-independent DPPs arising from
pickup coupling were found that had a significant emissive region
at the nuclear centre. The possibility that this points to an
$l$-dependent underlying DPP is supported by earlier~\cite{npa476}
model calculations. In these model calculations, $S_l$ calculated
for an explicitly $l$-dependent local potential were inverted leading
to an imaginary term with a strong emissive region at the nuclear centre.
Note that such an emissive region does not imply unitarity breaking:
an $l$-dependent potential can be devised for which $|S_l|\le 1$
for all $l$, and for which the $l$-independent equivalent potential
nevertheless has local emissive regions. Note that the emissive region
reported in Ref.~\cite{prc76} was in the DPP, not the full potential.

In Section~\ref{rmat}, it was reported that nucleon-$^4$He scattering
presented the following alternative: the potential exhibited either
\emph{waviness} or \emph{parity dependence}. RGM calculations
clearly imply a preference for parity dependence. Exchange
processes that occur with much heavier
scattering pairs of nuclei are also believed to lead to parity
dependence. Michel and Reidemeister~\cite{zp333} presented
strong evidence that the $^4$He - $^{20}$Ne interaction at 54.1
MeV contained a Majorana (i.e. $ (-1)^l$) term. Cooper and
Mackintosh~\cite{zp337} applied IP inversion to the $S_l$
derived from the potential of Ref.~\cite{zp333} and found a
parity-independent potential giving the same $S_l$. It had
a substantial oscillatory feature, suggesting that the
`waviness or parity-dependence' alternative is a general feature.
In this case, a quite small Majorana term led to quite a
considerably wavy $l$-independent equivalent.
This is not an argument against parity dependence, but, apart
from showing the power of IP inversion, it also suggests that
wavy potentials found in model independent OM fitting should be
considered seriously as a clue to underlying parity dependence.
Therefore, seeking perfect fits to elastic scattering data,
even when wavy potentials result, should not be dismissed as
`fitting elephants' (see page 223 of Ref.~\cite{brandan}); all
the information content of the experimental elastic scattering
data can be given meaning  --- that is surely desirable.

\section{Summary and outlook}
For a wide range of energies and projectile-target combinations, $S
\to V$ inversion, and also observable $\to V$ inversion, are
straightforward, and would be routine if the possible applications
were more widely appreciated. This review has concentrated on giving
some account of the information concerning nuclear interactions that
has been obtained, and left to other reviews the task of an
exposition of the mathematical basis of inverse scattering and the
attendant formal problems.

We note here some present limitations to the application of inversion,
with the hope that others might rise to the challenge of solving them:\\[5 mm]
{\bf Limitation 1: Spin.} At present successful inversions are
routine for the following cases: spin-zero on spin-zero, spin-\hf\
on spin zero; spin-1 on spin-zero. Spin-1 on spin-zero is currently
limited to determining the $T_{\rm R}$ interaction, the best
established tensor interaction. This limitation on spin gets in the
way of desired calculations mostly for the inversion of S-matrix
elements from theory rather than for S-matrices extracted from
experiments. For example, the calculation of the DPP for $^7$Li or
$^9$Be breakup cannot exploit the full S-matrix from a CC
calculation in which the spin of the projectile was treated
properly, even for a spin-zero target. The multiplicity of possible
tensor forms is discouraging, to say the least. However, a suitable
weighted S-matrix, such as that defined in Eqn.~\ref{sbar}, can
determine at least the central potential implicit in the S-matrix
output from a coupled channel calculation with particles of
spin $> \hf$ in the elastic channel. This approach was used for
deuterons in Ref.~\cite{prc77} and for $^6$Li (1+), $^7$Be (\hfm{3})
and $^8$B (2$^+$) in Ref.~\cite{npa834}, but no
spin-orbit potentials could be extracted.

Useful inversions can be performed in cases where both projectiles
have spin: in Ref.~\cite{npa592} separate potentials (without
spin-orbit interaction) were derived for each of the two values of
channel-spin for p + $^3$He and n + $^6$Li scattering (in both
cases, the Majorana term was very different for each value of
channel spin.) However, full inversion for higher spin remains
a challenge.\\[5 mm]
{\bf Limitation 2: Coupled channel inversion.} The full and
practical solution to the problem
\begin{equation} S^{J\pi}_{\alpha \alpha'} \to V_{\alpha \alpha'}(r)
\label{ccinv}\end{equation} remains elusive, though there have been
proposed extensions of the NS method. The one fully successful
coupled channel inversion procedure is the IP extension described
above leading to the non-diagonal $T_{\rm R}$ interaction for spin-1
projectiles. This suggests that further extensions are possible,
although the profusion of possible non-diagonal potentials is
challenging. Coupled channel inversion might make it possible to
answer questions such as: how do coupled reaction channels~\cite{npa209}
modify the deformation parameters that emerge in the analysis of inelastic
scattering from deformed nuclei?

Finally, a personal viewpoint: understanding the nucleon-nucleus
interaction potential is of fundamental importance in nuclear physics.
Much progress has been made in achieving a unified perspective
for positive and negative energies. Nevertheless, some form of local density
approximation is implicit in most studies and the
extracted potentials are local and $l$-independent. The non-locality
due to knock-on exchange is included in an approximate way, but the role
of other forms of non-locality, which are known to be present, is obscure
at best. Moreover, there is little understanding of how the interaction 
depends upon the particular last-occupied orbitals or collectivity of 
individual nuclei. Coupled channel-plus-inversion offers scope for
understanding the processes which occur when a nucleon, and particles
in channels coupled to the nucleon channels, interact with the curved nuclear
surface, with its density gradients.

\end{document}